\documentclass{article}

\usepackage{graphicx}
\usepackage{pifont}
\usepackage{latexsym}
\usepackage{amsfonts}
\usepackage{amsmath}
\usepackage{amssymb}
\newcommand{\shortcite}[1]{\cite{#1}}

\newcommand{\p}{{\rm P}}
\newcommand{\np}{{\rm NP}}

\newcommand{\elec}{\ensuremath{{\mathcal{E}}}}
\newcommand{\thetatwo}{\ensuremath{\Theta_2^p}}

\newcommand{\deltatwo}{\ensuremath{\Delta_2^{p}}}

\newtheorem{theorem}{Theorem}

\newtheorem{mylemma}[theorem]{Lemma}
\newcommand\qedblob{\ding{113}}
\def\literalqed{{\ \nolinebreak\hfill\mbox{\qedblob\quad}}}

\newtheorem{observation}[theorem]{Observation}

\newenvironment{proofs}{\noindent{\bf Proof.}\hspace*{1em}}{\literalqed\bigskip}

\showhyphens{}

\newcommand{\prob}[3]{
\begin{description}
  \item[Name:] #1
  \item[Given:] #2
  \item[Question:] #3
\end{description}}

\title{High-Multiplicity Election Problems\thanks{The journal version of this
paper~\cite{fit-hem:j:high-multiplicity} appears in the Journal of Autonomous Agents and Multi-Agent Systems, and 
the conference version of this paper~\cite{fit-hem:c:high-multiplicity} appears in
the Proceedings of the 17th International Conference on Autonomous Agents and Multiagent Systems.}}

\author{Zack Fitzsimmons \\
Dept.\ of Math.\ and Computer Science\\
College of the Holy Cross\\
Worcester, MA 01610 \and
Edith Hemaspaandra \\
Department of Computer Science\\
Rochester Institute of Technology \\
Rochester, NY 14623}%

\date{June 24, 2021}

\begin{document}
\sloppy

\maketitle

\begin{abstract}
The computational study of elections generally assumes that the preferences of the electorate come in as a list of votes. Depending on the context, it may be much more natural to represent the list succinctly, as the distinct votes of the electorate and their counts, i.e., high-multiplicity representation. We consider how this representation affects the complexity of election problems. High-multiplicity representation may be exponentially smaller than standard representation, and so many polynomial-time algorithms for election problems in standard representation become exponential-time. Surprisingly, for polynomial-time election problems, we are often able to either adapt the same approach or provide new algorithms to show that these problems remain polynomial-time in the high-multiplicity case; this is in sharp contrast to the case where each voter has a weight, where the complexity usually increases. In the process we explore the relationship between high-multiplicity scheduling and manipulation of high-multiplicity elections. And we show that for any fixed set of job lengths, high-multiplicity scheduling on uniform parallel machines is in P, which was previously known for only two job lengths.
We did not find any natural case where a polynomial-time election problem does not remain in P when moving to high-multiplicity representation. However, we found one natural NP-hard election problem where the complexity does increase, namely
winner determination for Kemeny elections.

\end{abstract}

\section{Introduction}\label{sec:intro}

Elections are an important and widely used tool for determining an outcome given the
preferences of several agents (see, e.g.,~\cite{bra-con-end-lan-pro:b:comsoc-handbook}).
In most of the computational studies
on elections, the preferences of the voters are represented
as a list of votes. Though this may be a reasonable representation
for paper ballots in political elections, in artificial
intelligence applications a more succinct representation where
the preferences of the electorate are represented as a list
of distinct votes and their counts may be more natural. For example,
this representation is used by the online preference repository
\textsc{PrefLib} for election data~\cite{mat-wal:c:preflib}.
Following terminology from %
scheduling
introduced by
Hochbaum and Shamir~\shortcite{hoc-sha:j:high-multiplicity}
we refer to this as high-multiplicity representation.

We consider how high-multiplicity representation can
affect the complexity of different election problems, and contrast
this with the case of weighted voters.

High-multiplicity representation
may be exponentially smaller than standard representation, and so
many polynomial-time algorithms for election problems in standard
representation become exponential-time. Surprisingly, we find that 
the complexity
of polynomial-time election problems generally remains in \p.
We explain this phenomenon by showing that many common proof
techniques that show that election problems are in \p\ can be
adapted to the high-multiplicity case~(see Section~\ref{sec:adapt}).

We also explore the relationship between the well-studied problem of
high-multiplicity scheduling
and high-multiplicity election manipulation
(see Section~\ref{s:scheduling}).
In the process, we show that for any fixed set of job lengths,
high-multiplicity scheduling on uniform parallel machines
(this is the case where different machines can have different
speeds) is in \p\ (Theorem~\ref{t:mdd}).
Previous work showed that for two job lengths,
high-multiplicity scheduling on uniform parallel machines
is in \p~\cite{mcc-sma-spi:j:two-lengths}, and very recent
work shows that for any fixed number of job lengths,
high-multiplicity scheduling on
{\em identical} parallel 
machines (basically bin packing)
is in \p~\cite{goe-rot:c:bin-packing-constant}.

In general, we find that the complexity of polynomial-time election problems in standard representation
does not increase when moving to the high-multiplicity case.
This is in line with related work that includes results for high-multiplicity elections 
(there simply called succinct elections), which
does not include any case where the complexity
increases~\cite{fal-hem-hem:j:bribery,fal-hem-hem-rot:j:llull,fal-hem-hem-rot:j:single-peaked-preferences,hem-hem-rot:j:hybrid,rus:t:borda}.
And this general
behavior is also found in the literature on high-multiplicity scheduling. For example, with respect
to scheduling,
Clifford and Posner~\shortcite{cli-pos:j:parallel-machine-scheduling} states:
``\ldots we do not know of a problem that
can be solved in polynomial time, under standard encoding, yet is \np-complete under
HM encoding.''

However, this does not mean that it is not possible for the complexity to
increase. For example, consider the following election system ExactlyHalfApproval.
Every voter votes by an approval vector
(i.e., states an approval or disapproval for each candidate),
and a candidate is a winner if they receive exactly half of all approvals.
Though simple, this election system is not particularly natural.
It has been chosen to show that an increase in complexity is possible.
Consider the problem of constructive control by adding candidates (CCAC)
for election system ExactlyHalfApproval.
Given an election in high-multiplicity representation,
a set of unregistered candidates, and a preferred candidate $p$, it is
\np-complete to determine whether $p$ can be made a winner of the election
using system ExactlyHalfApproval by adding unregistered candidates to the election.
This is because this problem is basically
subset sum, which is in \p\ for unary numbers by dynamic programming
and is \np-complete for
binary numbers~\cite{gar-joh:b:int}. (See Theorem~\ref{thm:jump} for a
straightforward proof of this example. Also, we mention in passing that ExactlyHalfApproval-Weighted-CCAC
is also \np-complete.)

We did find a natural case where the complexity increases, but for an \np-hard
election problem. We find that determining the winner of a Kemeny election, which is well-known to be $\thetatwo$-complete for standard representation~\cite{hem-spa-vog:j:kemeny}, is $\deltatwo$-complete for high-multiplicity representation.
This solves an open problem
from the previous work by Hemaspaandra, Spakowski, and Vogel~\shortcite{hem-spa-vog:j:kemeny}. (See Section~\ref{sec:kemeny}.)

\section{Preliminaries}\label{sec:prelim}

An {\em election} (in {\em standard representation})
is defined as a finite set of candidates
$C$, and a list of voters $V$, where each voter $v \in V$
is specified by their
preference order over the candidates.
In high-multiplicity representation, 
$V$ is not a list of voters, but instead
a list of distinct votes $v$ (preference orders) and their positive
integer counts $\kappa(v)$.
To see how these two definitions compare,
consider the election that in standard representation
consists of $C = \{a,b,c\}$ and
the following four voters: $(a > b > c)$,
$(a > b > c)$, $(a > b > c)$, and $(b > c > a)$.
This list of four voters in high-multiplicity representation is: 
$3 \times (a > b > c)$ and $1 \times (b > c > a)$.

We will compare complexity results between high-multiplicity
and weighted election problems.
In a {\em weighted election}, $V$ is still a list of voters, but now each
$v \in V$ has an associated positive integer weight $\omega(v)$, and
can be thought of as a coalition of $\omega(v)$ voters all voting the
same. It is important to note that weights are indivisible,
while the counts in high-multiplicity representation are not.\footnote{Xia,
Conitzer, and Procaccia~\shortcite{xia-con-pro:c:scheduling-manipulation} have a notion of ``divisible''
weights, but they allow noninteger divisions and so this is very different
from high-multiplicity. It is interesting that their paper reduces manipulation for this
notion to a scheduling problem, though a different scheduling problem than what we use.}
We assume that the preference order of each voter is a total order, 
i.e., he or she strictly ranks each candidate from most to least 
preferred.

An election system $\mathcal{E}$ maps an election to a set of
winners, where the winners can be any subset of the candidate set.
(This is known as the nonunique winner model.)
The problem \elec-Winner is defined below.
 
\prob{\elec-Winner}
{An election $(C,V)$ and a candidate $p \in C$.}
{Is $p$ a winner of $(C,V)$ using election system \elec?}

In the high-multiplicity case, \elec-High-Multiplicity-Winner, 
we use high-multiplicity representation for $V$,
and in the weighted case, \elec-Weighted-Winner, each
voter has a corresponding positive integer weight.

A scoring vector $\langle \alpha_1, \alpha_2, \dots, \alpha_m \rangle$,
$\alpha_i \ge \alpha_{i+1}$, 
defines an election system over $m$
candidates. Each candidate receives $\alpha_i$ points for each
vote where they are ranked $i$th, and the candidate(s) with the highest
score win.

A (polynomial-time uniform) {\em pure scoring rule} defines a family of scoring
vectors where
the $m$-candidate scoring vector can be computed in polynomial time in $m$,
and
the $(m+1)$-candidate scoring vector can be obtained from the
$m$-candidate scoring vector by adding a single
coefficient~\cite{bet-dor:j:possible-winner-dichotomy}.

\subsection{Manipulative Actions}

We examine the complexity of the following manipulative actions
on elections. For each problem, we present the definition for
the standard representation, and describe how the high-multiplicity and
weighted cases differ.

Manipulation is the most widely studied manipulative action on
elections.
The computational study of manipulation was introduced
by Bartholdi, Tovey, and Trick~\shortcite{bar-tov-tri:j:manipulating},
and was later extended for the coalitional, destructive, and weighted
cases by Conitzer, Sandholm, and
Lang~\shortcite{con-lan-san:j:when-hard-to-manipulate}.
We present the definition of %
constructive unweighted coalitional manipulation (CUCM) below.

\prob{\elec-CUCM}
{A set of candidates $C$, a list of nonmanipulative voters
$V_1$, a list
of $k$ manipulators $V_2$,\footnote{This can also be thought
of as specifying the number of manipulators, $k$, in unary.} and a preferred candidate $p \in C$.}
{Is there a way to set the preferences of the manipulators so that $p$
is a winner of the election $(C,V_1 \cup V_2)$ using election system~\elec?}

The weighted case (CWCM) is essentially the same %
as above, except
each of the voters
(both the nonmanipulators and the manipulators)
have an associated positive integer weight.
In the corresponding high-multiplicity case, we use high-multiplicity
representation for the nonmanipulators as well as for the manipulators,
which are represented as $k$ encoded in binary.

Electoral control denotes the family of manipulative actions that
consider an agent with control over the structure of the election,
called the election chair, who wants to ensure his or her preferred
outcome~\cite{bar-tov-tri:j:control}.
Below we define constructive control by adding voters (CCAV), which
is a very natural case of control and can be thought of as modeling
get-out-the-vote drives.

\prob{\elec-CCAV}
{A set of candidates $C$, a list of registered voters $V$,
a list of unregistered voters $U$, an add limit
$k \in \mathbb{N}$, and a preferred candidate $p \in C$.}
{Does there exist a list $U'$ of unregistered voters
$U' \subseteq U$ such that $U'$ consists of at most $k$ unregistered
voters and $p$ is a winner of $(C,V \cup U')$ using election system~\elec?}

In the standard model of weighted voter control
the parameter $k$ denotes the number of weighted voters the chair can
add/delete~\cite{fal-hem-hem:j:weighted-control}. In the
high-multiplicity case,
the only change from the \elec-CCAV definition above is that
we use high-multiplicity representation for 
both lists of voters.

\subsection{Computational Complexity}

We assume that the reader is familiar with the complexity classes
\p\ and \np . We also have results concerning the
complexity classes \thetatwo\ and \deltatwo .
\deltatwo\ denotes
the class of problems solvable in
\p\ with access to an \np\ oracle. \thetatwo\ is a subset of \deltatwo
, where the \p-machine can ask a log number of queries to an \np\ oracle (which is
equivalent to the class of problems solvable by a \p-machine that can ask one round
of parallel queries to an \np\ oracle~\cite{hem:j:sky}).

Most of our polynomial-time results for high-multiplicity  problems 
modify the proof of the problem in standard representation,
but some results will use the well-known result due
to Lenstra, which shows that
even though solving an integer linear program is \np-complete
in the general case~\cite{kar:b:reducibilities,bor-tre:j:bounds-linear-diophantine}, it is in \p\ when the number of variables
is fixed~\cite{len:j:integer-fixed}. Informally, an integer linear program
is a system of linear inequalities with integer variables and
coefficients.

\section{Adapting Approaches}\label{sec:adapt}

The general theme of this paper is that even though we would expect %
the complexity of election problems to increase (with respect to the length of the input) when %
using high-multiplicity representation, we find that
in general such increases do not occur.
In this section
we will discuss several common approaches used to 
show that election problems are in \p\ (in standard representation),
and describe how they can be adapted for high-multiplicity representation,
sometimes straightforwardly
and sometimes in a more complicated way.

In particular, we show how greedy approaches, limited brute-forcing, 
network flow, and edge matching/cover techniques can 
be adapted. To showcase these adaptations, we will show that
the dichotomy result for constructive control by adding voters (CCAV)
for pure scoring rules~\cite{hem-hem-sch:c:dichotomy-one} holds
for the high-multiplicity case.

Assuming P $\neq$ NP,
the following pure scoring rules are asymptotically
(i.e., for a large enough number of candidates)
the only cases where CCAV is in $\p$, both for standard and
high-multiplicity representation.

\begin{itemize}
 \item $\langle \alpha, \beta, 0, \ldots, 0 \rangle$, where
 $\alpha > \beta$.
   \item $t$-Approval: $\langle 1^t, 0, \dots, 0 \rangle$, where $t \le 3$.
  \item $t$-Veto: $\langle 1, \dots, 1, 0^t \rangle$, where $t \le 2$.
 \item $\langle 2, 1, \dots, 1, 0 \rangle$.
\end{itemize}

In contrast, all weighted cases are NP-complete,
except triviality (i.e., a scoring rule where every candidate scores the same), %
$1$-approval (plurality), $2$-approval, 
and $1$-veto~\cite{fal-hem-hem:j:weighted-control,lin:thesis:elections}.

The following is a case where we need to adapt a greedy approach
as well as to handle limited brute-forcing.

\begin{theorem}\label{thm:dichotwo}
For $\alpha > \beta \geq 0$,
$\langle \alpha,\beta,0, \dots, 0 \rangle$-CCAV is in \p\ for
high-multiplicity elections. %
\end{theorem}

\def\adaptdichotwo{
\begin{proofs}
We follow the proof of the claim from~\cite{hem-hem-sch:c:dichotomy-one}, which
is found in its corresponding technical report~\cite{hem-hem-sch:t:dichotomy-one},
and show how to adapt this to the high-multiplicity case.

The proof shows that there is a constant $\ell$ such that
it is never better to add $\ell$ votes that put $p$ second than
it is to add $\ell$ different votes that put $p$ first
(assuming that votes that rank the first two candidates
identically also rank all other candidates identically).
Let $V_1$ be the set of unregistered voters that put $p$ first and let
$V_2$ be the set of unregistered voters that put $p$ second.

So, if $p$ can be made a winner, $p$ can be made a winner by adding
at most $\ell-1$ voters from $V_2$ or in a way 
that does not leave $\ell$ different votes in $V_1$
unused.
    
In the first case, we can brute-force over all sets of $V_2$ voters
of size $< \ell$ (since $\ell$ is a constant).
Add these voters.
We are then left with the problem of checking
whether we can add at most $k'$ $V_1$ voters to an election to make
$p$ a winner, where $k'$ is $k$ minus the number of $V_2$ voters added.
In Hemaspaandra, Hemaspaandra, and Schnoor~\shortcite{hem-hem-sch:c:dichotomy-one},
this is done
by brute-forcing over every $j$ in $\{0, \ldots, k'\}$ and checking
whether we can add $j$ voters from $V_1$ to make $p$ a winner.
If we know $j$, we know what the score of $p$ after control will be,
and so the last part can then be done greedily by, for each candidate $a$,
adding as many voters that put $a$ second as possible
one voter at the time (until we have added $j$ voters total).

We have two problems in the high-multiplicity case.
The first is that we cannot in polynomial time
brute-force over all possible values of $j$, since $k'$ is not
polynomially bounded. The second problem is that we don't have the
time to add voters one at the time.  The second problem can of
course easily be fixed by, for each candidate $a$, adding as many voters
with $a$ in second place as possible all at once:
If $s_a$ is the initial score of $a$ and $\mbox{\it fs}_p$ is the final score
of $p$ (note that $\mbox{\it fs}_p = s_p + j \alpha$), we
can add at most  $\lfloor (\mbox{\it fs}_p - s_a)/\beta \rfloor$ voters with $a$ second.

To handle the first problem, note that we can 
formulate the
problem as an integer linear program with one variable ($j$), 
namely there is a $j$, $0 \leq j \leq k'$, such that
\[\sum_{a \in C - \{p\}} \min(\lfloor (s_p + j\alpha - s_a)/\beta \rfloor, v_a) \geq j,\]
where $v_a$ is the number of voters in $V_1$ that rank $a$ second.\footnote{Notice that the
ILP above checks that if we add the maximum number of voters such
that the score of each of the non-$p$ candidates is at most $s_p + j\alpha$, then we
have added at least $j$ voters. This implies that we can add $j$ voters in such a way that
$p$ is still a winner.}
This integer linear program can be solved in polynomial
time by Lenstra~\cite{len:j:integer-fixed}.\footnote{Note that
the ``$\min$'s'' here are not independent of each other. We can first sort the candidates so that when we
look at the successive ``$\min$'' values, these will always be the right-hand side for the first part and then the
left-hand side for the rest of the sequence.
This gives an ``OR'' of $\|C\|$ ILPs with a fixed number of variables. Also, note that we don't really
need an ILP approach in this case, but we will need that for the second case.}

It remains to show that the second case, where 
$p$ can be made a winner in a way that
does not leave $\ell$ different votes in $V_1$
unused, can be determined in polynomial time in the high-multiplicity case.

For the standard case, in
Hemaspaandra, Hemaspaandra, and Schnoor~\shortcite{hem-hem-sch:c:dichotomy-one}
this case is handled
by brute-forcing over all sets $S$ of at most $\ell - 1$ candidates
who are ranked second by unused $V_1$ voters, and then brute-forcing
over all possible sets of unused $V_1$ voters consistent with $S$.
We can describe a set of voters consistent with $S$
as a function $u: S \rightarrow \{1, \ldots, \|V_1\|\}$ such that $u(c)$ is the number
of unused $V_1$ voters that rank $c$ second.
We can brute-force over all such functions $u$ in polynomial time.
This way, we loop over all possible sets of unused $V_1$ voters, and so we also loop over
all possible sets of added $V_1$ voters.
 After adding the
 $V_1$ voters, we need to see
 if $p$ can be made a winner by adding $V_2$ voters. This can be done
 in a similar way as adding the $V_1$ voters in the first case above.

In the high-multiplicity case, we can still loop over all sets $S$ of 
at most $\ell - 1$ candidates. We can not brute-force all
functions $u$, but this will turn again into an integer linear
program, this time with $\|S\|$ variables (for the $u$ values),
$\|S\|$ variables for the $V_2$ voters voting for candidates in $S$,
and one
more variable to handle the remaining $V_2$ voters similarly to how we handled
the $V_1$ voters in the first case above.
This integer linear program can be solved in polynomial
time by Lenstra~\cite{len:j:integer-fixed}.~\end{proofs}}

\adaptdichotwo

We now show how edge matching/cover techniques can be adapted.

\begin{theorem}
CCAV is in \p\ for $t$-approval when $t \le 3$ and for $t$-veto
when $t \le 2$ for high-multiplicity elections.
\end{theorem}

\begin{proofs}
The results for $1$-approval, $2$-approval, and $1$-veto follow via
simple greedy algorithms that can easily be adapted to work in the 
high-multiplicity case.

$3$-approval-CCAV was shown to be in P by
Lin~\shortcite{lin:thesis:elections} using a
reduction to Simple $b$-Edge Matching for Multigraphs
(see~\cite{sch:b:combinatorial-opt}).
The essence of the reduction is that every
unregistered voter of the form $(\{p,x,y\} > \cdots)$ 
(by which we mean a voter who gives a point to $p$, $x$, and $y$)
corresponds to an edge $(x,y)$
in the constructed multigraph. For every candidate $c \neq p$, 
$c$ is a vertex in the graph and $b(c)$ is 
the final score of $p$ minus the initial score of $c$ (i.e.,
the number of points that we can add to $c$ while
keeping $c$'s score less than or equal to $p$'s score).
In the high-multiplicity case, we would have too many edges in the graph.
However, Capacitated $b$-Edge Matching (for graphs where edges 
have integer capacities) is also in P
(see~\cite{sch:b:combinatorial-opt}).
So, we simply
set the capacity of edge $(x,y)$ equal to the number of
unregistered voters of the form $(\{p,x,y\} > \cdots)$, and we let $b(c)$ be 
the final score of $p$ minus the initial score of $c$ as previously.

Similarly, $2$-veto-CCAV was shown to be in P by
reduction to Simple $b$-Edge Cover for Multigraphs.
In that reduction, every 
unregistered voter of the form $(\cdots > \{x, y\})$
corresponds to an edge 
$(x,y)$ in the constructed multigraph.  Again, we can modify this construction
to a reduction to Capacitated $b$-Edge Cover, which is also in P, by letting
the capacity of edge $(x,y)$ be the number of unregistered
voters of the form $(\cdots > \{x ,y\})$.~\end{proofs}

Another common technique to show that election problems are in
\p\ is network flow.

\begin{theorem}
$\langle 2, 1, \ldots, 1, 0 \rangle$-CCAV is in \p\ for high-multiplicity
elections.
\end{theorem}

\begin{proofs}
The essence of the standard representation proof is to (after some easy
preprocessing)
build a min-cost flow network. The capacities of the edges are
the scores of the candidates from the registered voters and
the multiplicities of the votes of the unregistered voters.
In the high-multiplicity case, we can use the exact same network.
The capacities are now binary integers, but min-cost network flow is
still in P in that case.~\end{proofs}

\noindent
{\bf Dynamic Programming}
There is one other common technique that is used to prove that election
problems are in P. This is dynamic programming. As alluded to
in the introduction,
dynamic programming approaches do not generalize to the high-multiplicity case.
Let's for example look at the result by Hemaspaandra and
Schnoor~\shortcite{hem-sch:c:dichotomy-two} that shows that for all pure 
scoring rules
with a constant number of different coefficients, manipulation is
in \p. This is shown by dynamic programming. And this algorithm
is not in \p\ for the high-multiplicity case, not even for very restricted versions
of this problem.
In the next section we consider a different approach to this problem using scheduling.

\section{Manipulation as Scheduling}\label{s:scheduling}

We now turn to the problem of manipulation for high-multiplicity elections
for pure scoring rules.
What does our election manipulation  problem have to do with scheduling?
As an example, consider scoring rule $\langle 4,3,3,2,0, \ldots, 0\rangle$
and let $\{p, c_1, \ldots, c_m\}$ be the set of candidates.
Let $s_c$ be the score of candidate $c$ from the nonmanipulators,
and let $k$ be the number of manipulators.  Note that we can
assume that all manipulators rank $p$ first, so that $p$'s final
score is $s_p + 4k$. We need to see if we can set the manipulators
such that every candidate $c \neq p$ scores at most $s_p + 4k$.
As also pointed out in Bachrach et al.~\shortcite{bac-lev-lew-zic:c:district-voting}
for the case without nonmanipulators,
we can view this as the following scheduling 
problem: We have $2k$ jobs of length 3 and $k$ jobs of length 2,
we have $m$ machines, machine $i$ corresponds to candidate $c_i$,
and machine $i$ has deadline $s_p + 4k - s_{c_i}$.
It is easy to see that if $p$ can be made a winner, then
we can schedule the jobs such that every machine meets its
deadline.
However, we note that the converse does not hold: 
We need to make sure that every machine has at most $k$ jobs
scheduled on it. We also need to be careful that
every voter ranks
each candidate exactly once. The construction from Hemaspaandra and
Schnoor~\shortcite{hem-sch:t:dichotomy-two} shows that
there is a successful manipulation if and only
if there is a successful schedule such that each machine
has at most $k$ jobs scheduled on it. Note that this also
gives the obvious equivalence for the corresponding high-multiplicity
versions of these problems.

More generally, we can state the following equivalence, which is
implicit in the work by Hemaspaandra and Schnoor~\shortcite{hem-sch:t:dichotomy-two},
which shows that manipulation is in \p\ for all pure scoring rules with a
constant number of different coefficients in standard representation.

\begin{theorem}
\label{t:man-sch}
Let $\langle \alpha_0, \alpha_1^{m_1}, \alpha_2^{m_2}, \ldots,
\alpha_{C}^{m_C}, 0^{m_{C+1}} \rangle$ be a scoring rule
such that $\alpha_0 \geq \alpha_1 > \alpha_2 > \cdots > \alpha_{C} > 0$
and $m_1, \ldots, m_{C+1}$ are positive integers
such that $m_1 + \cdots + m_{C+1} = m$. Let the set of candidates be
$\{p,c_1, \ldots, c_m\}$ and let there be $k$ manipulators.

Then $p$ can be made a winner if and only if
for $1 \leq j \leq C$, we can schedule $km_j$ jobs of length $\alpha_j$
on $m$ machines 
such that the $i$th machine has deadline $s_p + \alpha_0k - s_{c_i}$
\textbf{with the additional restriction that we schedule at most
$k$ jobs per machine}.
\end{theorem}

Since our problem is closely related to scheduling and
since high-multiplicity scheduling has been
well-studied, it makes sense to see what is known there, so that we
can see how easy or hard these problems are and so that we can
try to adapt the results.

First some terminology. It is easy to see that scheduling with machine-dependent
deadlines is equivalent to the well-known scheduling problem called
scheduling on uniform parallel machines (where different machines can
have different speeds)~\cite{mcc-sma-spi:j:two-lengths}.
In the high-multiplicity version of the scheduling with machine-dependent
deadlines 
problem, the input is given as $\ell_1, \ldots, \ell_C, n_1, \ldots, n_C, 
D_1, \ldots, D_m$, meaning that there are 
$n_i$ jobs of length $\ell_i$, and $m$ machines with 
deadlines $D_1, \ldots, D_m$.  If all deadlines are the same, 
the problem is equivalent to scheduling on identical parallel machines
(this problem is also known as bin packing and as the cutting-stock problem).
In the high-multiplicity version of that
problem, the input is given as
$\ell_1, \ldots, \ell_C, n_1, \ldots, n_C, m, D$ (note that this makes
the input even more succinct, and so a polynomial-time algorithm
for scheduling with machine-dependent deadlines
does not necessarily imply a polynomial-time algorithm
for the same problem where all machines have the same deadline).

In many contexts, the number of job lengths or item types is small and so
high-multiplicity scheduling for a constant number of item types
is an important problem.
Surprisingly, the complexity of this problem was open for a long time.
Leung~\cite{leu:j:scheduling-restricted} first studied this problem and provided a
pseudopolynomial algorithm for the case where all machines have the same deadline
that is polynomial in the length of the input and the number of items.
Only in 2001 was this
determined to be in P for two job lengths, both in the case
where all machines have the same deadline
as well as
the case with machine-dependent deadlines~\cite{mcc-sma-spi:j:two-lengths}.
The case for any constant number of job lengths 
was only very recently shown to be in P for the case where all machines
have the same deadline~\cite{goe-rot:c:bin-packing-constant}.
However, the algorithm from~\cite{goe-rot:c:bin-packing-constant} does not give a polynomial-time result
for the case of machine-dependent deadlines.

Given all this, do we have any chance of solving our high-multiplicity
manipulation problems? There are some glimmers of hope. The first
one is that the restriction to having at most $k$ jobs per machine
could make the problem easier (it could also make it harder of course).
Secondly, we know that the number of jobs of each length is
a multiple of $k$ (this could make the problem simpler).
Finally, we can take $\ell_1, \ldots, \ell_C$ to be constant, since
for a pure scoring rule with a constant number of different coefficients,
there are only a fixed number of coefficients that occur for any 
number of candidates, and by Theorem~\ref{t:man-sch}, the occurring coefficients
are exactly the occurring job lengths.
We will first show that for every fixed $C, \ell_1, \ldots, \ell_C$, 
high-multiplicity scheduling with machine-dependent deadlines
(and thus also high-multiplicity scheduling on uniform parallel machines)
is in P.
Note that this is a very natural problem, since in many
contexts, the set of job lengths or item types is fixed.
So this is a very interesting result in its own right.

\begin{theorem}
\label{t:mdd}
For any fixed set of job lengths,
high-multiplicity scheduling on uniform parallel machines
is in \p.\footnote{After the publication of our journal version~\cite{fit-hem:j:high-multiplicity}, we were informed that this result follows from
earlier work by Knop and Kouteck{\'y}~\cite{kno-kou:t:n-fold-integer-programming,kno-kou:j:n-fold-integer-programming}.}
\end{theorem}

\begin{proofs}
Let $C$ be the number of job lengths and let 
$\ell_1, \ldots, \ell_C$ be these job lengths.
As mentioned previously, 
high-multiplicity scheduling on uniform parallel machines
is equivalent to 
high-multiplicity scheduling on machines with machine-dependent
deadlines~\cite{mcc-sma-spi:j:two-lengths},
and this is the problem that we will show to be in P.

Given $n_1, \ldots, n_C$, where $n_i$ is the number of jobs
of length $\ell_i$, and $D_1, \ldots, D_m$, where $D_i$ is the
deadline for machine $i$,  we can
solve our scheduling problem in polynomial time, and pretty easily at that.

\smallskip

\noindent
{\bf Example Case.}
We will first explain the
core argument by looking at $C = 2$, $\ell_1 = 2$, and $\ell_2 = 3$
and after that we will explain how the argument generalizes.
So, we have $n_1$ jobs of length 2, $n_2$ jobs of length 3, and
$m$ machines with deadlines $D_1, \ldots, D_m$. And the question
is whether there exist nonnegative integers $x_{i}$
(the number of jobs of length 2 scheduled
on machine $i$) and $y_i$ (the number of jobs of length 3 scheduled
on machine $i$) such that $2x_{i} + 3y_{i} \leq D_i$,
$\sum_{i=1}^m x_{i} = n_1$, and $\sum_{i = 1}^m y_{i} = n_2$.

Consider the $i$th machine. We can group the jobs as ``sixes,'' by which we mean 3 jobs of length 2 or 2 jobs of length 3, and ``the leftovers,'' which consist of
0, 1, or 2 jobs of length 2 and 0 or 1 jobs of length 3. Note that the
sum of the leftovers is at most 7.

Let $D'_i$ be the largest integer that is $\leq D_i - 7$ 
and divisible by 6. Then $D_i'$ time can be scheduled with just sixes and
no leftovers. And so we can schedule successfully if and only if
there exist $a_1$ jobs of length
2 and $a_2$ jobs of length 3 that can be scheduled on $m$
machines with deadlines 
$D_1 - D'_1, D_2 - D'_2, \dots, D_m - D'_m$ such that the remaining
($n_1 - a_1$) jobs of length 2 and the remaining ($n_2 - a_2$)  jobs
of length 3 can be scheduled on $m$ machines with deadlines
$D'_1, \ldots, D'_m$ in sixes, i.e., as groups of 3 jobs of length 2
and groups of 2 jobs of length 3. 

Why is this in P? Note that $D_i - D'_i \leq 12$, since
$D'_i$ is the largest integer that is $\leq D_i - 7$ 
and divisible by 6, and so
$a_1 \leq 6m$ and $a_2 \leq 4m$, and we can use dynamic programming to
compute all values of $a_1$ and $a_2$ such that
$a_1$ jobs of length 2 and $a_2$ jobs of length 3
fit into $D_1 - D'_1, D_2 - D'_2, \dots, D_m - D'_m$.
Finally, check that 
$n_1 - a_1$ is divisible by 3, that $n_2 - a_2$ is divisible by 2, and
that $(n_1 - a_1)/3 + (n_2 - a_2)/2 \leq (D'_1 + \cdots + D'_m) / 6$.

This takes time $O(m^3 \log n)$, where $n$ is the number of jobs.
For the dynamic programming simply let,
for all  $a_1 \leq 6m$, $a_2 \leq 4m$, and $\widehat{m} \leq m$, 
$T[a_1,a_2,\widehat{m}] = 1$  if and only if $a_1$ jobs of length 2 and
$a_2$ jobs of length 3 fit into
$D_1 - D'_1, D_2 - D'_2, \dots, D_{\widehat{m}} - D'_{\widehat{m}}$.
There are $O(m^3)$ values to compute and it takes
$O(\log n)$ to compute one value, since all numbers involved are length $O(\log n)$
and there are a constant number of ways
to fit jobs into $D_i - D'_i$ (since $D_i - D'_i \leq 12$).

\smallskip

\noindent
{\bf General Case.} It is easy to see how the approach for
the example case generalizes to the case for each fixed set of
job lengths $\ell_1, \ldots, \ell_C$.
Let $\ell = lcm(\ell_1, \ldots, \ell_C)$.
Our input is $n_1, \ldots, n_C, D_1, \ldots, D_m$.
Consider the $i$th machine. We can group the jobs as ``$\ell$s,''
by which we mean $\ell/\ell_j$ jobs of length $\ell_j$
for some $j$,
and ``the leftovers,'' which consist of
$< \ell/\ell_j$ jobs of length $\ell_j$ for each $j$.
Note that the sum of the leftovers is at most $\sum_{j=1}^C (\ell/\ell_j - 1)\ell_j = \sum_{j=1}^C (\ell - \ell_j)$.

For each $i$, let $D'_i$ be the largest integer that is
$\leq D_i - \sum_{j=1}^C (\ell - \ell_j)$ and divisible by $\ell$.
Then $D_i'$ time can be scheduled with just $\ell$s and
no leftovers. And so we can schedule successfully if and only if
there exist $a_j$ jobs of length $\ell_j$, for $1 \leq j \leq C$,
that can be scheduled on $m$ machines with deadlines 
$D_1 - D'_1, D_2 - D'_2, \dots, D_m - D'_m$ such that all remaining
($n_j - a_j$) jobs of length $\ell_j$, for $1 \leq j \leq C$,
can be scheduled on $m$ machines with deadlines
$D'_1, \ldots, D'_m$ in $\ell$s, i.e., as groups of
$\ell/\ell_j$ jobs of length $\ell_j$.

Why is this in P? Note that
$D_i - D'_i \leq \sum_{j=1}^C (\ell - \ell_j) + \ell - 1$,
since $D'_i$ is the largest integer that is 
$\leq D_i - \sum_{j=1}^C (\ell - \ell_j)$ and divisible by $\ell$.
Since $C$ and $\ell_1, \ldots, \ell_C$ are fixed,
$a_j = O(m)$ and we can use dynamic programming to
compute all values of $a_1, \ldots, a_C$ such that
$a_j$ jobs of length $\ell_j$
fit into $D_1 - D'_1, D_2 - D'_2, \dots, D_m - D'_m$.
Finally, check that 
$n_j - a_j$ is divisible by $\ell_j$
and
that $\sum_{j=1}^C (n_j - a_j)/\ell_j \leq (D'_1 + \cdots + D'_m) / \ell$.

Note that since the $\ell_i$s and $C$ are constant,
the argument provided for the example case also shows
that the general case is in P.~\end{proofs}

Note that the above does not give a polynomial-time algorithm if only
$C$ is fixed and $\ell_1, \ldots, \ell_C$ are given as part of the input.
And so the above does not imply the result that for two job lengths,
high-multiplicity scheduling with machine-dependent deadlines is
in P from McCormick, Smallwood, and Spieksma~\shortcite{mcc-sma-spi:j:two-lengths}. 

However, as mentioned previously, the case where 
$\ell_1, \ldots, \ell_C$ are fixed is very natural, and so
this is an interesting result for scheduling.
But recall from Theorem~\ref{t:man-sch} that to solve manipulation,
we in addition have to ensure that there are at most $k$ jobs scheduled
on each machine. As mentioned earlier,
such a restriction can potentially make the complexity harder
(as well as easier).

What are the problems with adding the restriction that we
have at most $k$ jobs per machine in the
argument of the proof of Theorem~\ref{t:mdd}? Let's look
at the $C = 2, \ell_1 = 2, \ell_2 = 3$ case.
In the case without the restriction that we have at  most $k$ jobs
per machine,
all deadlines with the same value mod 6 are treated the same, and every
``six'' is treated the same (no matter what machine it's on).
This is no longer the case if we also require that every
machine handles at most $k$ jobs, since it then matters
if you schedule three jobs of length 2 or two jobs of length 3 in
a six.

However, we can show a number of high-multiplicity manipulation cases
to be in \p\ using scheduling, which is in contrast to the weighted case where only
triviality and plurality are in \p.
Note that the theorem below does not cover all of
the polynomial-time cases of manipulation for pure scoring rules in standard
representation, since
for example, we do not cover $\langle 2^{m/3}, 1^{m/3}, 0^{m/3} \rangle$
(though cases with at most two different coefficients can be handled easily).
Also note that there is a limit to how far this can be generalized,
since even in standard representation manipulation for Borda is NP-complete~\cite{bet-nie-woe:c:borda-manip,dav-kat-nar-wal:c:borda-manip}.

\begin{theorem}\label{thm:const}
    Let $\alpha_0 \geq \alpha_1 > \alpha_2 > \cdots > \alpha_{C} > 0$. %
Manipulation for pure scoring rule
$\langle \alpha_0, \alpha_1^{m_1}, \alpha_2^{m_2}, \ldots,
\alpha_{C}^{m_C}, 0^{m-(m_1 + \cdots + m_C)} \rangle$ 
is in \p\ for high-multiplicity elections.
\end{theorem}

\begin{proofs}
By Theorem~\ref{t:man-sch}, it suffices to show that the following problem
is in P: Given $k, D_1, \ldots, D_m$, can we
schedule $km_j$ jobs of length $\alpha_j$ on 
$m$ machines with deadlines $D_1, \ldots, D_m$ such that 
every machine meets its deadline and such that there are at most
$k$ jobs scheduled on each machine.

\smallskip

\noindent
{\bf Example Case.}
We first explain the core argument for $C = 2$, $\alpha_1 = 3$, $\alpha_2 = 2$,
$m_1 = 1$, $m_2 = 1$.
So, we are given
$k, D_1, \ldots, D_m$, and we are asking whether we can we schedule $k$ jobs
of length 2 and $k$ jobs of length 3 on
$m$ machines with deadlines $D_1, \ldots, D_m$ such that 
every machine meets its deadline and such that there are at most
$k$ jobs scheduled on each machine.
That is, we are asking 
whether there exist nonnegative integers
$x_{i}$ (the number of jobs of length 2 scheduled
on machine $i$)
and $y_i$ (the number of jobs of length 3 scheduled
on machine $i$)
such that $2x_{i} + 3y_{i} \leq D_i$,
$\sum_{i=1}^m x_{i} = k$, $\sum_{i = 1}^m y_{i} = k$, and
$x_{i} + y_{i} \leq k$.

Note that we can assume that $D_i \leq 3k$.
Also note that if $D_i \leq 2k$,
our new requirement that machine $i$ has at most 
$k$ jobs, i.e., $x_{i} + y_{i} \leq k$, follows from 
the old requirement that $2x_{i} + 3y_{i} \leq D_i$.

If $D_i \leq 2k$ for all $i$, then we can solve our
problem by using the algorithm from Theorem~\ref{t:mdd}.
If there are more than two machines with deadline $> 2k$, we can
schedule $k$ jobs of length 2 on machine 1,
$\lceil k/2 \rceil$ jobs of length 3 on machine 2,
and
$\lfloor k/2 \rfloor$ jobs of length 3 on machine 3.

It remains therefore to look at the case where there are one or
two machines with deadline $> 2k$. Without loss of generality, 
assume that the deadlines are in nonincreasing order and
that we have at least two machines. Then
our case is solvable if and only if there exist
$x_{1}, y_{1}, x_{2}, y_{2}$ such that
$2x_{1} + 3y_{1} \leq D_1$,
$2x_{2} + 3y_{2} \leq D_2$,
$x_{1} + y_{1} \leq k$, and
$x_{2} + y_{2} \leq k$,
such that 
$k - x_{1} - x_{2}$ jobs of length 2 and
$k - y_{1} - y_{2}$ jobs of length 3
can be scheduled on $m-2$ machines with
deadlines $D_3, \ldots, D_m$ (since all these deadlines are
$\leq 2k$, the requirement that we schedule at most $k$ jobs
on each machine will be automatically satisfied). We would
like to use the algorithm from Theorem~\ref{t:mdd}, but we would need
to run that algorithm for all values of
$x_{1}, y_{1}, x_{2}, y_{2}$, which is not polynomial.
The solution is to rephrase the algorithm from
Theorem~\ref{t:mdd} as an integer linear program with a fixed %
number of variables (see Lemma~\ref{l:ilp} below),
which we then combine with
the equations involving $x_{1}, y_{1}, x_{2}, y_{2}$ above.
This ILP can be solved in polynomial
time by Lenstra~\shortcite{len:j:integer-fixed}.

\smallskip

\noindent
{\bf General Case.}
Now let's consider how to generalize the approach used in
the example case to the general case.
Note that we can assume that $D_i \leq \alpha_1k$.
Also note that if $D_i \leq \alpha_Ck$,
our new requirement that machine $i$ has at most
$k$ jobs follows from the old requirement that
$D_i$ meets its deadline.

Let $\widehat{m}$ be a constant such that if there are
more than $\widehat{m}$ machines with deadline
$>\alpha_Ck$, we can greedily schedule all
jobs on these $\widehat{m}$ machines.
(Since we can always schedule $\lfloor (\alpha_Ck + 1) / \alpha_1 \rfloor$ jobs
on such a machine, such a constant exists.)

It remains therefore to look at the case where there are
at most $\widehat{m}$ machines with deadline $> \alpha_Ck$.
Without loss of generality,
assume that the deadlines are in nonincreasing order and
that we have at least $\widehat{m}$ machines.
Our case is solvable if and only if there exist
$x_{i,j}$ for $1 \leq i \leq \widehat{m}$ and $1 \leq j \leq C$,
the number of jobs scheduled on machine $i$ of length $\alpha_j$,
such that
$\sum_{j = 1}^C \alpha_j x_{i,j} \leq D_i$,
$\sum_{j = 1}^C x_{i,j} \leq k$, and such that
$k m_j - \sum_{i = 1}^{\widehat{m}} x_{i,j}$ jobs of length $\alpha_j$
can be scheduled on $m-\widehat{m}$ machines with
deadlines $D_{\widehat{m} + 1}, \ldots, D_m$ (since all these deadlines are
$\leq \alpha_C k$, the requirement that we schedule at most $k$ jobs
on each machine will be automatically satisfied).
Use the ILP formulation of the algorithm from
Theorem~\ref{t:mdd} (see Lemma~\ref{l:ilp} below)
to solve this last requirement and combine this
with our equations for
$x_{i,j}$ for $1 \leq i \leq \widehat{m}$ and $1 \leq j \leq C$,
to obtain an ILP with a fixed number of variables, which
can be solved in polynomial time by Lenstra~\shortcite{len:j:integer-fixed}.~\end{proofs}

It remains to show that the algorithm from Theorem~\ref{t:mdd}
can be rephrased as an integer linear program with a fixed number
of variables.

\begin{mylemma}
\label{l:ilp}
For any fixed set of job lengths, there is an ILP
with a fixed number of variables
that solves high-multiplicity scheduling on
parallel machines with machine-dependent deadlines.
\end{mylemma}

\begin{proofs}
Let $C$ be the number of job lengths and let 
$\ell_1, \ldots, \ell_C$ be these job lengths.
We are given $n_1, \ldots, n_C$, where $n_i$ is the number of jobs
of length $\ell_i$, and $D_1, \ldots, D_m$, where $D_i$ is the
deadline for machine $i$.
In the proof of
Theorem~\ref{t:mdd}, we proved that this scheduling problem is in P
using dynamic programming.
However, for the proof of
Theorem~\ref{thm:const} we need a different approach. 
Below we present
an ILP formulation with a fixed number of variables that computes if
we can schedule successfully that replaces the dynamic programming.
Again, to show the essence of the argument, we consider the
case of $C=2$, $\ell_1 = 2$, and $\ell_2 = 3$.

Look at a proof of Theorem~\ref{t:mdd} (the example case).
We are already pretty close:
We show that there exist $a_1$ and $a_2$ such that
$a_1$ jobs of length 2 and $a_2$ jobs of length 3
fit into $D_1 - D'_1, D_2 - D'_2, \dots, D_m - D'_m$,
$n_1 - a_1$ is divisible by 3, $n_2 - a_2$ is divisible by 2, and
$(n_1 - a_1)/3 + (n_2 - a_2)/2 \leq (D'_1 + \cdots + D'_m) / 6$.

So all that is left to do is to write 
``$a_1$ jobs of length 2 and $a_2$ jobs of length 3
fit into $D_1 - D'_1, D_2 - D'_2, \dots, D_m - D'_m$''
as an integer linear program with a fixed number of variables.
Recall that $D_i - D'_i \leq 12$. 
For $0 \leq j \leq 12$, let $d_j$ be the number of machines
$i$ such that $D_i - D_i' = j$. 

Our integer linear program formulation is as follows.
We introduce variables $a_{j,(r,t)}$ for
$0 \leq j \leq 12$ and $2r + 3t \leq j$.
$a_{j,(r,t)}$ stands for the number of machines $i$ such that $D_i - D'_i = j$
and such that $D_i - D'_i$ is scheduled with $r$ 2s and $t$ 3s.
For every $j$, $0 \leq j \leq 12$, we have an equation
$\sum_{2r + 3t \leq j} a_{j,(r,t)} = d_j$ and we
add the following obvious equations:
\[\sum_{0 \leq j \leq 12, 2r + 3t \leq j} a_{j,(r,t)} r = a_1.\]
\[\sum_{0 \leq j \leq 12, 2r + 3t \leq j} a_{j,(r,t)} t = a_2.\]
It is easy to see that the example case generalizes to the general
case, using the general case of the proof of Theorem~\ref{t:mdd}~\end{proofs}

\section{Fixed Numbers of Candidates}

It is reasonable to assume that the number of candidates in
an election may be fixed. For weighted voters
many problems are hard, even %
when the number of candidates is fixed.

\begin{theorem}
\begin{enumerate}
\item
$m$-candidate $\alpha$-CWCM is \np-complete for every
scoring rule $\alpha$ that is not plurality or
triviality~\cite{hem-hem:j:dichotomy}.
\item
$m$-candidate $\alpha$-CCAV and $m$-candidate $\alpha$-CCDV are each \np-complete for
every scoring rule $\alpha = \langle \alpha_1, \dots, \alpha_m \rangle$,
when $\|\{\alpha_1, \dots, \alpha_m \}\|
\ge 3$, for weighted elections~\cite{fal-hem-hem:j:weighted-control}.
\end{enumerate}
\end{theorem}

Faliszewski, Hemaspaandra, and Hemaspaandra~\shortcite{fal-hem-hem:j:bribery} showed
that high-multiplicity manipulation for scoring rules is in \p\ for any
fixed number of candidates by describing an ILP
with a fixed number of variables.

\begin{theorem}\cite{fal-hem-hem:j:bribery}
$m$-candidate $\alpha$-CUCM is in \p\ for every scoring rule
$\alpha$, for high-multiplicity elections.
\end{theorem}

A similar approach
can be used to show that high-multiplicity constructive control
by adding/deleting voters is in \p\ for every scoring rule for a
fixed number of candidates. %

\begin{theorem}\label{thm:ccav-fixed}
$m$-candidate $\alpha$-CCAV and
$m$-candidate $\alpha$-CCDV are each in \p\ for
every scoring rule
$\alpha = \langle \alpha_1, \dots, \alpha_m \rangle$, for
high-multiplicity elections.
\end{theorem}

\begin{proofs}
Given a set of candidates $C$, a list of registered voters $V$
in high-multiplicity representation, a
list of unregistered voters $U$ in high-multiplicity representation,
an addition limit 
$k \in \mathbb{N}$, and a preferred candidate $p \in C$
we can determine if the chair can ensure that $p$ wins in polynomial
time. The argument closely follows
the approach used by Faliszewski, Hemaspaandra, and Hemaspaandra~\shortcite{fal-hem-hem:j:bribery} to show that
$m$-candidate $\alpha$-bribery and $m$-candidate $\alpha$-CUCM, for every scoring rule $\alpha$ are each in \p\ for  high-multiplicity (there called
succinct) elections.

We describe an integer linear program with a fixed number of variables,
which due to the result from Lenstra~\shortcite{len:j:integer-fixed}, can be solved in polynomial time.

Since the number of candidates $m$ is fixed, we know that only $m!$
different preference orders  are possible. We list these
preference orders in order in the following way:
$o_1, \dots, o_{m!}$.
For $1 \le i \le m!$, let $k_i$ be the number of voters in $V$ with preference order
$o_i$, and let $k'_i$ be the number of voters in $U$ with preference order $o_i$.
Let $x_1, \dots, x_{m!}$ be variables. The constants used as input to our linear program are the coefficients of the scoring vector
$\alpha_1, \dots, \alpha_m$, the counts
for the registered voters $k_1, \dots, k_{m!}$,
the counts for the unregistered voters
$k'_1, \dots, k'_{m!}$, and the addition limit $k$.
We have the following constraints.

We first need to ensure that all of our variables have a nonnegative
value. So, $\forall i, 1 \le i \le m!$, $$x_i \ge 0.$$

We also need to ensure that no more than the number of unregistered
votes of each type are added. So, $\forall i, 1 \le i \le m!$, $$x_i \le k'_i.$$

The following constraint ensures that no more than $k$ total votes 
are added. $$\sum_{i=1}^{m!} x_i \le k.$$

Our final constraint ensures that no other candidate has a score
greater than $p$. (Note that ${\rm pos}(c,j)$ below denotes the
position of candidate $c$ in preference order $o_j$.)
So, for all $c \in C-\{p\}$, %
$$\sum_{j=1}^{m!} \alpha_{{\rm pos}(p,j)}(k_j + x_j) \ge \sum_{j=1}^{m!} \alpha_{{\rm pos}(c,j)}(k_j + x_j).$$

It is easy to see that the above constraints are satisfied
if and only if it is possible to add at most $k$ unregistered 
voters from $U$ such that $p$ wins.

The above integer linear program can be easily modified for the case of
deleting voters.~\end{proofs}

\section{Natural Increase in Complexity}\label{sec:kemeny}

In general, we find that the complexity of polynomial-time election
problems in standard representation
does not increase when moving to the high-multiplicity case.
However, this does not mean that such an increase is not
possible.
Recall the election system ExactlyHalfApproval defined in the introduction.

\begin{theorem}\label{thm:jump}
ExactlyHalfApproval-CCAC is in \p\ and ExactlyHalfApproval-High-Multiplicity-CCAC is \np-complete.
\end{theorem}

\begin{proofs}
This problem is basically subset sum, which is in P for
unary numbers and NP-complete for binary numbers~\cite{gar-joh:b:int}.
To show that
ExactlyHalfApproval-CCAC is in P, let $s_1, \ldots, s_m$ be the approval scores
of candidates $c_1, \ldots, c_m$, let $c_1$ be the preferred candidate
and let $c_1, \ldots, c_\ell$ be the registered candidates. We want to
know if there exists a subset $I$ of $\{\ell+1, \ldots, m\}$
of size at most $k$ such that 
\[2s_1 =  \sum_{1 \leq i \leq \ell} s_i + \sum_{i \in I} s_i.\]
This can easily be computed in polynomial time, using dynamic programming.
Simply let, for all $\ell+1 \leq i \leq m$, $0 \leq k' \leq k$,
$0 \leq t \leq \sum_{1 \leq i \leq m} s_i$,
$S[i,t,k'] = 1$ if and only if there exists a size-$k'$ subset $I$
of $\{\ell+1, \ldots, m\}$ 
such that $\sum_{1 \leq i \leq \ell} s_i + \sum_{i \in I} s_i = t$.
This can be done in polynomial time, since there are only polynomially
many $i$, $k'$, and $t$ to consider.

When the input is given in high-multiplicity representation,
there are exponentially many values
for $t$ possible. In this case, we can show that it is NP-complete.
We reduce from Partition: Given $\{k_1,\ldots, k_m\}$ such that $\sum k_i = 2K$,
we need to determine if a subset of these integers sums to $K$. We have $m+1$ candidates,
preferred candidate $p$ and candidates $c_1,\ldots, c_m$. We have $K$ voters
that approve of only $p$, and $k_i$ voters that approve of only $c_i$. $p$ is
the only registered voter and the addition limit is $m$. It is immediate
that there is a subset of $k_1, \ldots, k_m$ that sums to $K$ if and
only if $p$ can be made a winner by adding candidates.~\end{proofs}

Note that this complexity increase also holds for the weighted case, since for
candidate control problems, the weighted and high-multiplicity cases coincide
(if their winner problems coincide, as they typically do).

As mentioned in the introduction, this system is not particularly natural.
We will now present a natural case where the complexity increases,
namely winner determination for Kemeny elections.

The Kemeny rule was introduced in~\cite{kem:j:no-numbers}, and has
several desirable properties~\cite{lev-you:j:condorcet}.
A candidate $p$ is a Kemeny winner if $p$ is ranked first in a 
Kemeny consensus, i.e., a linear order $>$ over the candidates
that minimizes the Kendall tau distance to $V$, i.e., that minimizes
$\sum_{a, b \in C, a > b} \|\{v \in V \ | \ b >_v a\}\|$,
where $>_v$ is the preference order of voter $v$.

\begin{observation}
Kemeny-Weighted-Winner is equivalent to Kemeny-High-Multiplicity-Winner.
\end{observation}

Notice that the above observation holds since we
do not modify the votes to score the election.
This is
not the case for Dodgson and Young elections, which
we discuss in Appendix~\ref{app:dod-you}. %
Kemeny-Winner was shown NP-hard in~\cite{bar-tov-tri:j:manipulating}
and $\thetatwo$-complete in~\cite{hem-spa-vog:j:kemeny}.
$\thetatwo$-hardness was shown by a chain
of three reductions, ultimately showing that
the $\thetatwo$-complete problem
Min-Card-Vertex-Cover-Compare reduces to Kemeny-Winner.

Footnote 3 in~\cite{hem-spa-vog:j:kemeny} points out that
Kemeny-High-Multiplicity-Winner is in $\deltatwo$
and explicitly 
leaves the exact complexity of this problem as an open question.
We will now show that Kemeny-High-Multiplicity-Winner
(and Kemeny-Weighted-Winner) are in fact $\deltatwo$-complete.
We define the following \deltatwo-complete weighted version
of Min-Card-Vertex-Cover-Compare.

\prob{Min-Weight-Vertex-Cover-Compare}
{Vertex-weighted graphs $G$ and $H$ such that
    $\omega(G) = \omega(H)$, where $\omega(G)$ denotes the weight of the graph.\footnote{Note that
the weight of a vertex-weighted graph is the sum of the weights of each of its vertices.}}
{Is the weight of $G$'s minimum-weight vertex cover
$\le$ the weight of $H$'s minimum-weight vertex cover?}

\begin{mylemma}
\label{l:ksw}
Min-Weight-Vertex-Cover-Compare polynomial-time many-one
reduces to Kemeny-High-Multiplicity-Winner.
\end{mylemma}

\begin{proofs}
This follows by careful inspection and modification of the proof
from Hemaspaandra, Spakowski, and Vogel~\cite{hem-spa-vog:j:kemeny}
that Min-Card-Vertex-Cover-Compare polynomial-time many-one reduces to Kemeny-Winner.
That proof consists of a chain of three reductions:
\begin{enumerate}
\item Min-Card-Vertex-Cover-Compare reduces to Vertex-Cover-Member.
\item Vertex-Cover-Member reduces to Feedback-Arc-Set-Member.
\item Feedback-Arc-Set-Member reduces to Kemeny-Winner.
\end{enumerate}

Since that proof consists of a chain of three
reductions between $\thetatwo$-complete problems, we need to define
suitable $\deltatwo$-complete weighted versions of the two
intermediate problems and show that the weighted versions of
the reductions still hold.
In essence, we ``lift'' the constructions and proofs
from $\thetatwo$ to $\deltatwo$.  This works surprisingly smoothly.

We define the following weighted versions of the two intermediate problems.

\prob{Weighted-Vertex-Cover-Member}
{Vertex-weighted graph $G$ and vertex $v$ in $G$.}
{Is $v$ a member of a minimum-weight vertex cover of $G$?}

\prob{Weighted-Feedback-Arc-Set-Member}
{Irreflexive and antisymmetric edge-weighted digraph $G$ and vertex $v$ in
$G$.}
{Is there a minimum-weight feedback arc set of $G$ that contains all
arcs entering $v$?}

And we show the following.
\begin{enumerate}
\item Min-Weight-Vertex-Cover-Compare reduces to
Weighted-Vertex-Cover-Member (Lemma~\ref{l:a1}).

\item Weighted-Vertex-Cover-Member reduces to Weighted-Feedback-Arc-Set-Member
(Lemma~\ref{l:a2}).

\item Weighted-Feedback-Arc-Set-Member reduces to Kemeny-High-Multiplicity-Winner
(Lemma~\ref{l:a3}).
\end{enumerate}
This implies that Min-Weight-Vertex-Cover-Compare reduces to
Kemeny-High-Multiplicity-Winner.~\end{proofs}

\begin{mylemma}
\label{l:a1}
Min-Weight-Vertex-Cover-Compare
polynomial-time many-one
reduces to
Weighted-Vertex-Cover-Member.
\end{mylemma}

\begin{proofs}
We extend the construction from Lemma 4.12 from~\cite{hem-spa-vog:j:kemeny}
to vertex-weighted graphs.
Given two vertex-weighted graphs $G$ and $H$ with $\omega(G) = \omega(H)$,
construct graph $F$ as follows. $F = (G \cup (\{v\},\emptyset))
+ (H \cup (\{w\},\emptyset))$,\footnote{For two disjoint graphs
$G_1$ and $G_2$, the join $G_1 + G_2$ is defined as follows:
$V(G_1 + G_2) = V(G_1) \cup V(G_2)$ and $E(G_1 + G_2) = E(G_1) \cup E(G_2) \cup
\{\{v,w\} \ | \ v \in V(G_1), w \in V(G_2)\}$.}
keeping the weights of the vertices
in $G$ and $H$, and letting $\omega(v) = \omega(w) = 1$. 
Map $(G,H)$ to $(F,w)$. Write $mvc$ for the minimum vertex cover weight of
a vertex-weighted graph. We extend the argument
from Lemma 4.12 from~\cite{hem-spa-vog:j:kemeny}
to show that this mapping reduces Min-Weight-Vertex-Cover-Compare
to Weighted-Vertex-Cover-Member.

Note that $V$ is a vertex cover of $F$ if and only if
\begin{enumerate}
\item $V$ contains all vertices in $G \cup (\{v\},\emptyset)$ and a 
vertex cover of $H$. The lightest such vertex covers have weight
$\omega(G) + 1 + mvc(H)$ and do not contain $w$.

\item $V$ contains all vertices in $H \cup (\{w\},\emptyset)$
and a vertex cover of $G$.  The lightest such vertex covers have weight
$mvc(G) + \omega(H) + 1$ and contain $w$.
\end{enumerate}
Since $\omega(G) = \omega(H)$ it is immediate that
$mvc(G) \leq mvc(H)$ if and only if there is a minimum-weight
vertex cover of $F$ that contains $w$.
\end{proofs}

\begin{mylemma}
\label{l:a2}
Weighted-Vertex-Cover-Member polynomial-time many-one
reduces to Weighted-Feedback-Arc-Set-Member.
\end{mylemma}

\begin{proofs}
We extend the construction from the proof of Lemma 4.8
from~\cite{hem-spa-vog:j:kemeny}, which itself is an extension
of the standard reduction from Vertex-Cover to Feedback-Arc-Set
from~\cite{kar:b:reducibilities},
to graphs with weights.  Given a vertex-weighted graph $G$, define
edge-weighted digraph $H = (W,A)$ as follows.
\begin{enumerate}
\item $W = \{v,v' \ | \ v \in V(G)\}$.
\item $A$ consists of the following edges.
\begin{enumerate}
\item For $v \in V(G)$, edge $(v,v')$ with weight $\omega(v)$.
\item For $\{v,w\} \in E(G)$,  edges
$(v',w)$ and $(w',v)$ with weight $\omega(G)$.
\end{enumerate}
\end{enumerate}
The reduction maps $(G,\hat{v})$ to $(H,\hat{v}')$.

If $V'$ is a vertex cover of $G$, then
by the proof of Lemma 4.9 from~\cite{hem-spa-vog:j:kemeny},
$\{(v,v') \ | \ v \in V'\}$ is a feedback arc set
for $H$.  Note that the weight of this feedback arc set is $\omega(V')$
and that if $V'$ contains $\hat{v}$, then
$\{(v,v') \ | \ v \in V'\}$ contains all arcs entering $\hat{v}'$.

Now suppose that $A'$ is a feedback arc set of
$H$. Again by the proof
of Lemma~4.9 from~\cite{hem-spa-vog:j:kemeny},
$V' = \{v \in V(G) \ | \ \exists w \in W:  (v,w) \in A' \mbox{ or }
(v',w) \in A'\}$ forms a vertex cover of $G$.
Note that $\omega(V') \leq \omega(A')$ (since
for every $v \in V'$, there exists a $w \in W$ such that
$(v,w) \in A'$ or $(v',w)$ in $A'$ and each such edge contributes
$\omega(v)$ or $\omega(G)$ to $\omega(A')$) and
that if $A'$ contains $(\hat{v},\hat{v}')$, then
$V'$ contains $\hat{v}$.
\end{proofs}

\begin{mylemma}
\label{l:a3}
Weighted-Feedback-Arc-Set-Member
polynomial-time many-one
reduces to Kemeny-High-Multiplicity-Winner.
\end{mylemma}

\begin{proofs}
We extend the construction from the proof of
Lemma 4.2 of~\cite{hem-spa-vog:j:kemeny} to graphs with weights.
Given an irreflexive and antisymmetric edge-weighted digraph
$G$ and vertex $v$ in $G$, we first multiply all edge weights by 2 (this
does not change the membership of $(G,v)$ in
Weighted-Feedback-Arc-Set-Member). We then use McGarvey's
construction~\cite{mcg:j:election-graph} to construct in polynomial
time an election in high-multiplicity representation
that has $G$ as its weighted majority graph.
The proof of Lemma 4.2 of~\cite{hem-spa-vog:j:kemeny} shows that
there is a minimum-weight feedback arc set of $G$ that contains all
arcs entering $v$ if and only if $v$ is a Kemeny winner of the
constructed election.
\end{proofs}

This completes the proof of Lemma~\ref{l:ksw}.
To prove the main theorem of this section it remains
to show the following lemma.

\begin{mylemma}
\label{l:wvc}
Min-Weight-Vertex-Cover-Compare is $\deltatwo$-hard.
\end{mylemma}

\begin{proofs}
For a formula $\phi(x_1, \ldots, x_n)$, we view an assignment for
$\phi$ as the $n$-bit string $\alpha_1 \cdots \alpha_n$ such that
$\alpha_i$ gives the assignment for variable $x_i$. We identify
$\alpha$ with the binary number (between 0 and $2^n - 1$)
that it represents.
\cite{hem-hem-rot:j:online} shows that it is $\deltatwo$-hard
to compare the maximal satisfying assignments of two cnf formulas.
By negating the variables in the formulas,
we obtain the following $\deltatwo$-hard promise problem, which we will 
reduce to Min-Weight-Vertex-Cover-Compare to prove Lemma~\ref{l:wvc}.

\prob{MINSATASG$_{\leq}$}
{Two satisfiable 3cnf formulas $\phi(x_1, \ldots, x_n)$ and
$\psi(x_1, \ldots, x_n)$.}
{Is the minimal satisfying assignment of $\phi$
$\leq$ the minimal satisfying assignment of $\psi$?}

Let $\phi(x_1, \ldots, x_n)$ and
$\psi(x_1, \ldots, x_n)$ be two satisfiable 3cnf formulas.
Without loss of generality, assume that $\phi$ and $\psi$
have the same number $m$ of clauses (simply pad).

Let $f(\phi)$ be the graph computed by the standard 
reduction from 3SAT to Vertex-Cover from~\cite{kar:b:reducibilities}.
Then $f(\phi)$ consists of $2n + 3m$ vertices: 
$\{x_i, \overline{x_i} \ | \ 1 \leq i \leq n\} \cup
\{a_i, b_i, c_i \ | \ 1 \leq i \leq m\}$, and the following edges:
$\{x_i, \overline{x_i}\}$ for $1 \leq i \leq n$,
$\{a_j, b_j\}, \{a_j, c_j\}$, and $\{b_j,c_j\}$ for $1 \leq j \leq m$,
and
if the $j$th clause in $\phi$ is $\ell_1 \vee \ell_2 \vee \ell_3$,
where $\ell_r \in \{x_i, \overline{x_i} \ | \ 1 \leq i \leq n\}$,
then we have edges $\{a_j,\ell_1\}$, $\{b_j,\ell_2\}$, $\{c_j,\ell_3\}$.

The properties of $f$ that we need here, which follow
immediately from the proof of~\cite{kar:b:reducibilities}, are as follows.
\begin{enumerate}
\item $f(\phi)$ does not have a vertex cover of size less than $n + 2m$.
\item If $W$ is a vertex cover of size $n + 2m$, then
$W \cap \{x_i, \overline{x_i} \ | \ 1 \leq i \leq n\}$
corresponds to a satisfying assignment of $\phi$, in the sense that
$\| W \cap \{x_i, \overline{x_i}\} \| = 1$ for $1 \leq i \leq n$, 
and $\alpha_1 \cdots \alpha_n$ defined as
$\alpha_i = 1$ if and only if $x_i \in W$ is a
satisfying assignment for $\phi$.
\item If $\alpha$ is a satisfying assignment for $\phi$, then
there is a vertex cover of size $n+2m$ such that
$W \cap \{x_i, \overline{x_i} \ | \ 1 \leq i \leq n\}$ corresponds to
this assignment, i.e., the set
$\{x_i \ | \ \alpha_i = 1\} \cup \{\overline{x_i} \ | \ \alpha_i = 0\}$
can be extended to a vertex cover of size $n + 2m$ by adding $2m$ vertices 
from $\{a_i, b_i, c_i \ | \ 1 \leq i \leq m\}$. 
\end{enumerate}

Now set the weights of the vertices as follows:
$\omega(a_i) = \omega(b_i) = \omega(c_i) = 2^n$,
$\omega(x_i) = 2^n + 2^{n-i}$, and
$\omega(\overline{x_i}) = 2^n$.
Note that for $W$ a set of vertices, $\|W\| = \lfloor \omega(W) / 2^n \rfloor$.
In particular, a minimum-weight vertex cover will also have
minimum size.

We will now show that $n$-bit string $\alpha$ is the
smallest satisfying assignment of $\phi$ if and only if
$f(\phi)$ has a minimum-weight vertex cover of weight $(n+2m)2^n + \alpha$.
(Recall that we
interpret $n$-bit string $\alpha$ as a binary number between $0$ and $2^n - 1$.)

If $\alpha$ is a satisfying assignment for $\phi$, then
by property 3 above
the set $\{x_i \ | \ \alpha_i = 1\}
\cup \{\overline{x_i} \ | \ \alpha_i = 0\}$
can be extended to a vertex cover of size $n + 2m$ by adding $2m$ vertices 
from $\{a_i, b_i, c_i \ | \ 1 \leq i \leq m\}$. The weight of this
vertex cover is $(n + 2m)2^n + \alpha$.  Since the minimum size of a
vertex cover is $n + 2m$, the weight of a minimum-weight vertex cover
is $(n + 2m)2^n + \beta$, for some $\beta$ such that
$0 \leq \beta < 2^n$. Such a $\beta$
corresponds to a satisfying assignment by property 2 above.

Since $\psi$ is also a satisfiable 3cnf formula over $x_1, \ldots, x_n$ 
with $m$ clauses, it also holds that $n$-bit string $\alpha$ is the
smallest satisfying assignment of $\psi$ if and  only if
$f(\psi)$ has a minimum weight vertex cover of weight $(n+2m)2^n + \alpha$.

This then implies that $\phi$'s minimal satisfying assignment is
$\leq$ $\psi$'s minimal satisfying assignment 
if and only if the weight of
$f(\phi)$'s minimum-weight vertex cover is 
less than or equal to the weight of
$f(\psi)$'s minimum-weight vertex cover.
Note that $\omega(f(\phi)) = \omega(f(\psi))$.
This completes the reduction from 
MINSATASG$_{\leq}$ to 
Min-Weight-Vertex-Cover-Compare.~\end{proofs}

\noindent
This completes the proof of the main theorem of this section.

\begin{theorem}
Kemeny-Weighted-Winner and  Kemeny-High-Multiplicity-Winner are
$\deltatwo$-complete.
\end{theorem}

Dwork et al.~\shortcite{dwo-kum-nao-siv:c:rank-aggregation} show that
Kemeny-Winner is already NP-hard for four voters. In fact, one can easily
combine the techniques from Hemaspaandra, Spakowski, and
Vogel~\shortcite{hem-spa-vog:j:kemeny} and
Dwork et al.~\shortcite{dwo-kum-nao-siv:c:rank-aggregation} to obtain the
following theorem.
For details, see the appendix.

\begin{theorem}
\label{t:kw4}
Kemeny-Winner for four voters is $\thetatwo$-complete.%
\end{theorem}

Do we get the same complexity jump
from $\thetatwo$-complete to $\deltatwo$-complete
if we look at the high-multiplicity (or weighted)
case for four votes?
Note that in the high-multiplicity case we allow many voters with the same
vote, and so the Kendall tau distance of a linear order
$>$ over the candidates to the set of voters can be large. 
However, even though the distances can be large, it is easy to see that
there are only a polynomial number of possibilities.
Namely, if the 
multiplicities of the four votes are $k_1$, $k_2$, $k_3$, and $k_4$,
the only possible distances are
$\ell_1 k_1 + \ell_2 k_2 + \ell_3 k_3 + \ell_4 k_4$, where
$0 \leq \ell_i \leq \|C\|(\|C\| - 1) / 2$.
Consider the following problem
\prob{Kemeny-High-Multiplicity-Score}
{An election $(C,V)$ in high-multiplicity representation,
candidate $c \in C$, an integer $k$.}
{Does there exist a linear order $>$ over the candidates
such that $c$ is ranked first and the Kendall tau distance
of $>$ to $V$ is at most $k$?}
It is easy to see that Kemeny-High-Multiplicity-Score is in NP.
Now for each candidate $c$,
simply query the Kemeny-High-Multiplicity-Score oracle for all possible distances
in parallel, and then determine whether $p$ is a winner.

\begin{theorem}
Kemeny-Weighted-Winner and Kemeny-High-Multiplicity-Winner
for four votes are $\thetatwo$-complete.
\end{theorem}

\section{Conclusions and Future Work}\label{sec:conclusion}

High-multiplicity representation is a very natural way to represent
an election. This representation may be exponentially
smaller than the standard representation and so we would expect
to see an increase in complexity (with respect to the length of the
input). However, we were able to either adapt the approaches used for 
standard representation or provide new
algorithms to show that polynomial-time election problems generally
remain in \p. We also explored the relationship between
high-multiplicity scheduling
and manipulation of high-multiplicity elections,
which led to a new result in scheduling.

There are several interesting directions for future work.
Is it possible for a high-multiplicity election problem to be harder than
the corresponding weighted problem?
Can we find natural cases, for election problems as well as for scheduling
problems, where the complexity increases from P
to NP-hard when going to high-multiplicity representation?

\bigskip

\noindent
{\bf Acknowledgments:} We thank Martin Kouteck{\'y},
Marc Neveling, Robin Weishaupt, and the
referees for their helpful comments and suggestions.
And we thank the AAAI-17 Student Abstract referees for their helpful comments
and suggestions on our preliminary work on this topic~\cite{fit-hem:c:succinct-student-sa}. 
This work was supported in part by
a National Science Foundation Graduate
Research Fellowship under NSF grant no.\ DGE-1102937.

\newcommand{\etalchar}[1]{$^{#1}$}

\appendix

\section{Appendix}

\subsection{Proof of Theorem~\ref{t:kw4}:
Kemeny-Winner for four voters is $\thetatwo$-complete.}

We combine the constructions from~\cite{hem-spa-vog:j:kemeny}
and~\cite{dwo-kum-nao-siv:c:rank-aggregation}, to reduce Feedback-Arc-Set-Member to
Kemeny-Winner for four voters. 

Given an irreflexive and antisymmetric digraph $G = (V,A)$ and vertex
$v \in V$,
we first compute graph $G'$ as done
in~\cite{dwo-kum-nao-siv:c:rank-aggregation}.
$G' = (V',A')$ such that $V' = V \cup A$ and
$A' = \{(v,(v,w)), ((v,w),w) \ | \ (v,w) \in A\}$.
Note that $G$ has a feedback arc set of size $k$ if and only if
$G'$ has a feedback arc set of size $k$. In addition, and important
for our proof, $(G,v)$ is in Feedback-Arc-Set-Member if and only
if $(G',v)$ is in Feedback-Arc-Set-Member. Also note that
$G'$ is irreflexive and antisymmetric.
Now apply the 
reduction from Feedback-Arc-Set-Member to
Kemeny-Winner from~\cite{hem-spa-vog:j:kemeny} on $(G',v)$. 
This reduction outputs an election $g(G')$ that corresponds to
$G'$ and $(G',v)$ is in Feedback-Arc-Set-Member if and only
if $(g(G'),v)$ is in Kemeny-Winner.  The construction
from~\cite{dwo-kum-nao-siv:c:rank-aggregation} shows that we need
only four voters in $g(G')$.

\subsection{Dodgson and Young Elections}\label{app:dod-you}

In Section~\ref{sec:kemeny} we consider the complexity of winner determination for
Kemeny elections.
In this section we consider the complexity of winner determination for
Dodgson~\shortcite{dod:unpubMAYBE:dodgson-voting-system} and for
Young~\shortcite{you:j:extending-condorcet}, which as with Kemeny, is 
$\thetatwo$-complete for standard representation~\cite{hem-hem-rot:j:dodgson,rot-spa-vog:j:young}.
The Dodgson score of a
candidate $c$ is the number of swaps between adjacent candidates needed in the preferences of
the voters in the election to make $c$ a Condorcet winner (a candidate that beats
each other candidate pairwise).  The Dodgson winner(s) of an election are the candidate(s) with
minimum score. The Young score of a candidate $c$ is the size of a
largest subcollection of voters for which $c$ is a weak Condorcet winner
(a candidate that beats-or-ties each other candidate pairwise). The Young winner(s)
of an election are the candidate(s) with maximum score.

For Young and Dodgson for weighted elections we can consider both the case where we charge
by the voter (e.g., for Dodgson the score is based on the number of swaps regardless of the weight of
the voters where candidates are swapped) and the case where we charge by the voter-weight
(e.g., for Dodgson a swap in the vote of a weight $\omega$ voter counts as $\omega$ swaps).
Clearly, by brute-force, we have the following.\footnote{Note that to compute the Dodgson
    score for a given candidate $c$ it suffices to
    consider only swaps that move $c$ up. So for each voter we only have to look
    at a linear number of possibilities (depending only on the position of $c$).}

\begin{theorem}
For a fixed number of voters, Young-Weighted-Winner
and Dodgson-Weighted-Winner are in \p, both 
in the model where we charge by the voter and
in the model where we charge by the voter-weight.
\end{theorem}

The high-multiplicity case for Young is also in P, but this seems to need much more
powerful machinery.

\begin{theorem}
For a fixed number of votes, Young-High-Multiplicity-Winner is in \p.
\end{theorem}

\begin{proofs}
Given an election $(C,V)$ in high-multiplicity representation and
a candidate $p$, we can determine if $p$ is a Young winner in the
following way.

First, consider the following ILP that determines if the Young score of a candidate
$c$ is $\ge$ a given $k$.
Let $v_1, \ldots, v_\ell$ be the distinct votes in $V$.
The ILP has variables $k_1, \dots, k_{\ell}$ that correspond
to the number of voters ``kept'' for each vote and
the following constraints.

\begin{align}
  \forall i, 1 \le i \le \ell, & \ \ k_i \ge 0 \label{young:ilp1} \\
  \forall i, 1 \le i \le \ell, & \ \ k_i \le \kappa(v_i) \label{young:ilp2} \\
                               & \ \ \sum_{i=1}^{\ell} k_i \ge k \label{young:ilp3} \\
    \forall a \neq c, & \ \ \sum_{1 \le i \le \ell, c >_i a} k_i \ge \sum_{1 \le i \le \ell, a >_i c} k_i \label{young:ilp4}
\end{align}

Constraints~\ref{young:ilp1}, \ref{young:ilp2}, and~\ref{young:ilp3} ensure that the number
of voters kept for each vote is consistent with the given election and that the total number of
voters kept is at least $k$.
Constraint~\ref{young:ilp4} ensures that $c$ is a weak Condorcet winner.

We can then determine if a candidate $p$ is a Young winner by separately computing the Young score of
each candidate using the ILP described above (which can be solved in polynomial time
by Lenstra~\cite{len:j:integer-fixed}) and binary search, and $p$ is a winner if and only if
$p$ has maximum Young score.~\end{proofs}

This case of a fixed number of votes is a natural case to consider,
since this models 
scenarios where all voters aligning with
the same political party or group may vote the same.
This is reminiscent to related work where the 
preferences of the voters are single-peaked
(and so only a restricted number
of the total possible votes occur) (see, e.g.,~\cite{fal-hem-hem-rot:j:single-peaked-preferences}),
but instead we restrict the number of votes in an
election without assuming structure among them.

For the general weighted cases where we charge by the voter, it is
easy to see that the scores are polynomially bounded, and so we
can state the following result.

\begin{theorem}
Young-Weighted-Winner and Dodgson-Weighted-Winner 
in the model where we charge by the voter are
$\thetatwo$-complete.
\end{theorem}

It is easy to see that 
Young-Weighted-Winner and Dodgson-Weighted-Winner 
in the model where we charge by voter-weight as well as
Young-High-Multiplicity-Winner and Dodgson-High-Multiplicity-Winner are in
$\deltatwo$, by using binary search to find the scores
of all candidates.

We conjecture that, as in the case for Kemeny, we can ``lift''
the $\thetatwo$-hardness proofs for Dodgson and Young
to $\deltatwo$-hardness proofs for the weighted cases.
We do not think that this ``lifting'' will work for the
high-multiplicity cases and leave the exact complexity of
these problems, which
we know to be $\thetatwo$-hard and in $\deltatwo$, as an
open question.


\begin{thebibliography}{DKNW11}

\bibitem[BCE{\etalchar{+}}16]{bra-con-end-lan-pro:b:comsoc-handbook}
F.~Brandt, V.~Conitzer, U.~Endriss, J.~Lang, and A.~Procaccia.
\newblock {\em Handbook of Computational Social Choice}.
\newblock Cambridge University Press, 2016.

\bibitem[BD10]{bet-dor:j:possible-winner-dichotomy}
N.~Betzler and B.~Dorn.
\newblock Towards a dichotomy of finding possible winners in elections based on
  scoring rules.
\newblock {\em Journal of Computer and System Sciences}, 76(8):812--836, 2010.

\bibitem[Bla58]{bla:b:polsci:committees-elections}
D.~Black.
\newblock {\em The Theory of Committees and Elections}.
\newblock Cambridge University Press, 1958.

\bibitem[BLLZ16]{bac-lev-lew-zic:c:district-voting}
Y.~Bachrach, O.~Lev, Y.~Lewenberg, and Y.~Zick.
\newblock Misrepresentation in district voting.
\newblock In {\em Proceedings of the 25th International Joint Conference on
  Artificial Intelligence}, pages 81--87. IJCAI/AAAI Press, July 2016.

\bibitem[BNW11]{bet-nie-woe:c:borda-manip}
N.~Betzler, R.~Niedermeier, and G.~Woeginger.
\newblock Unweighted coalitional manipulation under the {B}orda rule is
  {NP}-hard.
\newblock In {\em Proceedings of the 22nd International Joint Conference on
  Artificial Intelligence}, pages 55--60. {IJCAI/AAAI} Press, July 2011.

\bibitem[BT76]{bor-tre:j:bounds-linear-diophantine}
I.~Borosh and L.~Treybig.
\newblock Bounds on positive integral solutions of linear {D}iophantine
  equations.
\newblock {\em Proceedings of the American Mathematical Society},
  55(2):299--304, 1976.

\bibitem[BTT89]{bar-tov-tri:j:manipulating}
J.~{{Bartholdi}}, III, C.~Tovey, and M.~Trick.
\newblock The computational difficulty of manipulating an election.
\newblock {\em Social Choice and Welfare}, 6(3):227--241, 1989.

\bibitem[BTT92]{bar-tov-tri:j:control}
J.~{{Bartholdi}}, III, C.~Tovey, and M.~Trick.
\newblock How hard is it to control an election?
\newblock {\em Mathematical and Computer Modeling}, 16(8/9):27--40, 1992.

\bibitem[CP01]{cli-pos:j:parallel-machine-scheduling}
J.~Clifford and M.~Posner.
\newblock Parallel machine scheduling with high multiplicity.
\newblock {\em Mathematical Programming}, 89(3):359--383, 2001.

\bibitem[CSL07]{con-lan-san:j:when-hard-to-manipulate}
V.~Conitzer, T.~Sandholm, and J.~Lang.
\newblock When are elections with few candidates hard to manipulate?
\newblock {\em Journal of the ACM}, 54(3):Article~14, 2007.

\bibitem[DKNS01]{dwo-kum-nao-siv:c:rank-aggregation}
C.~Dwork, R.~Kumar, M.~Naor, and D.~Sivakumar.
\newblock Rank aggregation methods for the web.
\newblock In {\em Proceedings of the 10th International World Wide Web
  Conference}, pages 613--622. ACM Press, March 2001.

\bibitem[DKNW11]{dav-kat-nar-wal:c:borda-manip}
J.~Davies, G.~Katsirelos, N.~Narodytska, and T.~Walsh.
\newblock Complexity and algorithms for {B}orda manipulation.
\newblock In {\em Proceedings of the 25th AAAI Conference on Artificial
  Intelligence}, pages 657--662. AAAI Press, August 2011.

\bibitem[Dod76]{dod:unpubMAYBE:dodgson-voting-system}
C.~Dodgson.
\newblock A method of taking votes on more than two issues.
\newblock Pamphlet printed by the Clarendon Press, Oxford, and headed ``not yet
  published'' (see the discussions
  in~\protect\cite{mcl-urk:b:polsci:classics,bla:b:polsci:committees-elections},
  both of which reprint this paper), 1876.

\bibitem[FH17]{fit-hem:c:succinct-student-sa}
Z.~Fitzsimmons and E.~Hemaspaandra.
\newblock The complexity of succinct elections.
\newblock In {\em Proceedings of the 31st AAAI Conference on Artificial
  Intelligence (Student Abstract)}, pages 4921--4922. {AAAI} Press, February
  2017.

\bibitem[FH18]{fit-hem:c:high-multiplicity}
Z.~Fitzsimmons and E.~Hemaspaandra.
\newblock High-multiplicity election problems.
\newblock In {\em Proceedings of the 17th International Conference on
  Autonomous Agents and Multiagent Systems}, pages 1558--1566. IFAAMAS, July
  2018.

\bibitem[FH19]{fit-hem:j:high-multiplicity}
Z.~Fitzsimmons and E.~Hemaspaandra.
\newblock High-multiplicity election problems.
\newblock {\em Autonomous Agents and Multi-Agent Systems}, 33(4):383--402,
  2019.

\bibitem[FHH09]{fal-hem-hem:j:bribery}
P.~Faliszewski, E.~Hemaspaandra, and L.~Hemaspaandra.
\newblock How hard is bribery in elections?
\newblock {\em Journal of Artificial Intelligence Research}, 35:485--532, 2009.

\bibitem[FHH15]{fal-hem-hem:j:weighted-control}
P.~Faliszewski, E.~Hemaspaandra, and L.~Hemaspaandra.
\newblock Weighted electoral control.
\newblock {\em Journal of Artificial Intelligence Research}, 52:507--542, 2015.

\bibitem[FHHR09]{fal-hem-hem-rot:j:llull}
P.~Faliszewski, E.~Hemaspaandra, L.~Hemaspaandra, and J.~Rothe.
\newblock Llull and {Copeland} voting computationally resist bribery and
  constructive control.
\newblock {\em Journal of Artificial Intelligence Research}, 35:275--341, 2009.

\bibitem[FHHR11]{fal-hem-hem-rot:j:single-peaked-preferences}
P.~Faliszewski, E.~Hemaspaandra, L.~Hemaspaandra, and J.~Rothe.
\newblock The shield that never was: {Societies} with single-peaked preferences
  are more open to manipulation and control.
\newblock {\em Information and Computation}, 209(2):89--107, 2011.

\bibitem[GJ79]{gar-joh:b:int}
M.~Garey and D.~Johnson.
\newblock {\em Computers and Intractability: {A} Guide to the Theory of
  {NP}-Completeness}.
\newblock {W. H. Freeman and Company}, 1979.

\bibitem[GR14]{goe-rot:c:bin-packing-constant}
M.~Goemans and T.~Rothvo{\ss}.
\newblock Polynomiality for bin packing with a constant number of item types.
\newblock In {\em Proceedings of the 25th Annual ACM-SIAM Symposium on Discrete
  Algorithms}, pages 830--839. SIAM, January 2014.

\bibitem[Hem89]{hem:j:sky}
L.~Hemachandra.
\newblock The strong exponential hierarchy collapses.
\newblock {\em Journal of Computer and System Sciences}, 39(3):299--322, 1989.

\bibitem[HH07]{hem-hem:j:dichotomy}
E.~Hemaspaandra and L.~Hemaspaandra.
\newblock Dichotomy for voting systems.
\newblock {\em Journal of Computer and System Sciences}, 73(1):73--83, 2007.

\bibitem[HHR97]{hem-hem-rot:j:dodgson}
E.~Hemaspaandra, L.~Hemaspaandra, and J.~Rothe.
\newblock Exact analysis of {D}odgson elections: {L}ewis {C}arroll's 1876
  voting system is complete for parallel access to {NP}.
\newblock {\em Journal of the ACM}, 44(6):806--825, 1997.

\bibitem[HHR09]{hem-hem-rot:j:hybrid}
E.~Hemaspaandra, L.~Hemaspaandra, and J.~Rothe.
\newblock Hybrid elections broaden complexity-theoretic resistance to control.
\newblock {\em Mathematical Logic Quarterly}, 55(4):397--424, 2009.

\bibitem[HHR14]{hem-hem-rot:j:online}
E.~Hemaspaandra, L.~Hemaspaandra, and J.~Rothe.
\newblock The complexity of online manipulation of sequential elections.
\newblock {\em Journal of Computer and System Sciences}, 80(4):697--710, 2014.

\bibitem[HHS14a]{hem-hem-sch:c:dichotomy-one}
E.~Hemaspaandra, L.~Hemaspaandra, and H.~Schnoor.
\newblock A control dichotomy for pure scoring rules.
\newblock In {\em Proceedings of the 28th AAAI Conference on Artificial
  Intelligence}, pages 712--720. {AAAI} Press, July 2014.

\bibitem[HHS14b]{hem-hem-sch:t:dichotomy-one}
E.~Hemaspaandra, L.~Hemaspaandra, and H.~Schnoor.
\newblock A control dichotomy for pure scoring rules.
\newblock Technical Report arXiv:1404.4560~[cs.GT], arXiv.org, April 2014.

\bibitem[HS91]{hoc-sha:j:high-multiplicity}
D.~Hochbaum and R.~Shamir.
\newblock Strongly polynomial algorithms for the high multiplicity scheduling
  problem.
\newblock {\em Operations Research}, 39(4):648--653, 1991.

\bibitem[HS16a]{hem-sch:t:dichotomy-two}
E.~Hemaspaandra and H.~Schnoor.
\newblock Complexity dichotomies for unweighted scoring rules.
\newblock Technical Report arXiv:1604.05264~[cs.CC], arXiv.org, April 2016.

\bibitem[HS16b]{hem-sch:c:dichotomy-two}
E.~Hemaspaandra and H.~Schnoor.
\newblock Dichotomy for pure scoring rules under manipulative electoral
  actions.
\newblock In {\em Proceedings of the 22nd European Conference on Artificial
  Intelligence}, pages 1071--1079. {IOS} Press, August/September 2016.

\bibitem[HSV05]{hem-spa-vog:j:kemeny}
E.~Hemaspaandra, H.~Spakowski, and J.~Vogel.
\newblock The complexity of {Kemeny} elections.
\newblock {\em Theoretical Computer Science}, 349(3):382--391, 2005.

\bibitem[Kar72]{kar:b:reducibilities}
R.~Karp.
\newblock Reducibilities among combinatorial problems.
\newblock In R.~Miller and J.~Thatcher, editors, {\em Complexity of Computer
  Computations}, pages 85--103, 1972.

\bibitem[Kem59]{kem:j:no-numbers}
J.~Kemeny.
\newblock Mathematics without numbers.
\newblock {\em Daedalus}, 88:577--591, 1959.

\bibitem[KK16]{kno-kou:t:n-fold-integer-programming}
D.~Knop and M.~Kouteck{\'{y}}.
\newblock Scheduling meets $n$-fold integer programming.
\newblock Technical Report arXiv:1603.02611v1~[cs.DS], arXiv.org, March 2016.

\bibitem[KK18]{kno-kou:j:n-fold-integer-programming}
D.~Knop and M.~Kouteck{\'{y}}.
\newblock Scheduling meets $n$-fold integer programming.
\newblock {\em Journal of Scheduling}, 21(5):493--503, 2018.

\bibitem[{Len}83]{len:j:integer-fixed}
H.~{Lenstra, Jr.}
\newblock Integer programming with a fixed number of variables.
\newblock {\em Mathematics of Operations Research}, 8(4):538--548, 1983.

\bibitem[Leu82]{leu:j:scheduling-restricted}
J.~Leung.
\newblock On scheduling independent tasks with restricted execution times.
\newblock {\em Operations Research}, 30(1):163--171, 1982.

\bibitem[Lin12]{lin:thesis:elections}
A.~Lin.
\newblock {\em Solving Hard Problems in Election Systems}.
\newblock PhD thesis, Rochester Institute of Technology, Rochester, NY, 2012.

\bibitem[McG53]{mcg:j:election-graph}
D.~McGarvey.
\newblock A theorem on the construction of voting paradoxes.
\newblock {\em Econometrica}, 21(4):608--610, 1953.

\bibitem[MSS01]{mcc-sma-spi:j:two-lengths}
T.~McCormick, S.~Smallwood, and F.~Spieksma.
\newblock A polynomial algorithm for multiprocessor scheduling with two job
  lengths.
\newblock {\em Mathematics of Operations Research}, 26(1):31--49, 2001.

\bibitem[MU95]{mcl-urk:b:polsci:classics}
I.~McLean and A.~Urken.
\newblock {\em Classics of Social Choice}.
\newblock University of Michigan Press, 1995.

\bibitem[MW13]{mat-wal:c:preflib}
N.~Mattei and T.~Walsh.
\newblock \textsc{PrefLib}: A library for preferences.
\newblock In {\em Proceedings of the 3rd International Conference on
  Algorithmic Decision Theory}, pages 259--270, November 2013.

\bibitem[RSV03]{rot-spa-vog:j:young}
J.~Rothe, H.~Spakowski, and J.~Vogel.
\newblock Exact complexity of the winner problem for {Young} elections.
\newblock {\em Theory of Computing Systems}, 36(4):375--386, 2003.

\bibitem[Rus07]{rus:t:borda}
N.~Russell.
\newblock Complexity of control of {B}orda count elections.
\newblock Master's thesis, Rochester Institute of Technology, 2007.

\bibitem[Sch03]{sch:b:combinatorial-opt}
A.~Schrijver.
\newblock {\em Combinatorial Optimization: {P}olyhedra and Efficiency}.
\newblock Springer-Verlag, 2003.

\bibitem[XCP10]{xia-con-pro:c:scheduling-manipulation}
L.~Xia, V.~Conitzer, and A.~Procaccia.
\newblock A scheduling approach to coalitional manipulation.
\newblock In {\em Proceedings of the 11th ACM Conference on Electronic
  Commerce}, pages 275--284. ACM Press, June 2010.

\bibitem[YL78]{lev-you:j:condorcet}
H.~Young and A.~Levenglick.
\newblock A consistent extension of {Condorcet}'s election principle.
\newblock {\em SIAM Journal on Applied Mathematics}, 35(2):285--300, 1978.

\bibitem[You77]{you:j:extending-condorcet}
H.~Young.
\newblock Extending {Condorcet}'s rule.
\newblock {\em Journal of Economic Theory}, 16(2):335--353, 1977.

\end{thebibliography}
\end{document}